\documentclass[useAMS,usenatbib]{mn2e}
\usepackage{graphicx}
\usepackage{rotating}
\usepackage{longtable}
\usepackage{lscape}

\begin{document}
\title[Low z QSO properties]
{Low redshift  quasars in the SDSS Stripe 82: Associated companion galaxies  and signature of star formation. }

\author[Bettoni  et al.]{
	D. Bettoni$^{1}$\thanks{E--mail: {\tt daniela.bettoni@oapd.inaf.it}}, 
        R. Falomo$^{1}$,
	J.~K.~Kotilainen$^{2,3}$ and
	K. ~Karhunen$^{3}$  \\	
       	$^{1}$ INAF -- Osservatorio Astronomico di Padova, Vicolo dell'Osservatorio 5, I-35122 Padova (PD), Italy\\
       	$^{2}$ Finnish Centre for Astronomy with ESO (FINCA), University of Turku, V\"ais\"al\"antie 20, FI-21500 Piikki\"o, Finland\\
       	$^{3}$ Tuorla Observatory, Department of Physics and Astronomy, University of Turku, FI-21500 Piikkio, Finland.
	}
\maketitle
\label{firstpage}

\begin{abstract}

We  obtained optical spectroscopy of close ($<$ 80 kpc) companion objects of a sample of 12 low redshift quasars (z $<$ 0.3 ) selected from the SDSS Stripe82  area and that are in the subsample of 52 QSOs for which both multicolor host galaxies properties and galaxy environment was recently investigated in detail. 
 We found that for 8 out of 12 sources the companion galaxy is  associated to the QSO having a difference of radial velocity that is less than 400 km/s. Many of these associated companions exhibit  [OII] $\lambda$3727 \AA~ emission lines suggestive  of episodes of (recent) star formation possibly induced by  past interactions. The SFR of the companion galaxies as derived from [OII] line luminosity is, however,  modest, with a median value of 1.0 $\pm$0.8 $M_{\odot}$/yr, and the emission lines are barely consistent with expectation from gas ionization by the QSO. The role of the QSO for inducing star formation in close companion galaxies appears meager.
 For three objects we also detect  the starlight spectrum of the QSO host galaxy which is characterized by absorption lines of old stellar population and [OII] emission line.
 
\end{abstract}


\section{Introduction}

Active supermassive black holes (quasars, QSO) are rare objects in the Universe but they 
 represent a key ingredient to fully understand the processes that have built the galaxies. 
 In fact a general consensus has emerged in the last decade that all sufficiently massive galaxies have a massive black hole in their center and therefore have the possibility to shine as a quasar. In spite
of several studies aimed to understand the mechanisms that activate and fuel the active
nuclei of galaxies only fragmented data are available. The most accredited responsible
for transforming a dormant massive black hole into a luminous quasar remains dissipative
tidal interactions and galaxy merging. Minor and major merging events may have a key
role for triggering and fueling the nuclear/quasar activity. These effects strictly
depend on the global properties of the galaxy environment \citep[see e.g.][]{kauffmann,dimatteo}. 

At low-redshift, quasars follow the large-scale structure traced by galaxy clusters 
but they eschew the very centre of clusters \citep{soch02,soch04}. On the other hand, on 
small scales (projected distance $<$ 1 Mpc) the quasar environment appears overpopulated by blue disc galaxies having a strong star formation rate (SFR) \citep{coldwell03,coldwell06}. 
However at Mpc scale comparing the environments of quasars to those of galaxies has
given conflicting results. Early studies on scales of 10Mpc suggest that quasars are more
strongly clustered than galaxies \citep[see e.g.][]{shanks}, while later studies based on surveys such as the Two Degree Field (2dF) and the  SDSS have found that the galaxy densities of quasars and inactive galaxies are comparable  \citep[see e.g.][]{smith,wake,serber,matsuoka14}.  

Since early imaging studies of QSOs and their environments it was noted that in a number of cases companion galaxies were found close to the quasars  \citep{Stockton82,Hutchings92}.  Further investigations were carried out using narrow band images that were tuned to  detect strong emission line (mainly [OII] 3727\AA   ~and  [OIII] 4959,5007 \AA) companions at the rest frame redshift of the QSOs \citep[e.g.]{Hutchings93,Hutchings95a,Hutchings95b,Hutchings95c}. These observations of close companion galaxies are thus suggestive of a physical association with the quasars  \citep{Stockton87}.  
A more sound proof of the association between the QSO and a close companion galaxy comes  from the accurate  spectroscopic measurement of both their redshifts.   Examples of these association are shown by \citet{Hutchings92}  and became more evident by HST images (e.g. \cite{Bahcall97}  for low z objects and by high quality images obtained in the near-IR for high z sources \citep[e.g.]{Falomo01,Falomo08}. Recently further spectroscopic evidence of this was given by \citet{Villar} for a small sample of low redshift QSOs.
 
These companion galaxies of quasars could be the product of a major merger of galaxies that led to the nuclear activities of QSOs  \citep{Stockton82, Bekki99}. 
 It is therefore of interest to  investigate by spectroscopy these close companions  in order to prove their physical  association to the QSO and to search for signature of recent star formation. Either these issues can probe and eventually support  the hypothesis that merging of galaxies is linked to the fueling and triggering of powerful nuclear activity in galaxies.

We have recently carried out  an imaging  study of the close companions around $\sim$ 50 low redshift (z $<$ 0.3) quasars \citep{bettoni15}. The targets were  extracted from  a larger dataset of $\sim$ 400 quasars at z $<$ 0.5 for which both the host galaxies and their galaxy environments were studied  \citep{falomo14,karhunen14} based on  deep multicolor images of SDSS Stripe82  \citep{annis2011}.
 It was found that for about 60\% of the QSOs there is a companion galaxy at projected distance less than 50 kpc and for half of them there are two or more companions.
 The comparison with a sample of inactive galaxies of comparable luminosity and  at similar redshift of the QSO hosts yields analogous results suggesting a weak link between the presence of close companions and nuclear activity. However, we found  an  indication that bright  companion galaxies are a factor of 2  more frequent in QSO  than in inactive galaxies \citep{bettoni15}.

In order to better understand the connection between nuclear activity and interaction/merging phenomena we should identify which companion galaxies are physically connected to the QSO (same redshift) from those that are just projected.
Unfortunately only in few cases the redshift of these companions was available from SDSS spectroscopy and as expected  some of these companions are found to be  not physically associated with the QSO (the companions are  mainly foreground galaxies) thus hampering a sound  interpretation of the effect of close companion galaxies on the nuclear activity. As a further step in this study we have therefore obtained spectroscopy of the companion galaxies of the QSO in the above sample in order to probe their physical association and to search for signature of recent star formation.

The paper is organized as follows: in Section 2 we present our QSO sample. Section 3 describes the analysis of the data and the main properties of the host galaxies and in Section 4 we discuss our results and we compare our findings with those of \citet{matsuoka14}. We adopt the concordance cosmology with H$_0$ = 70 km s$^{-1}$ Mpc$^{-1}$, $\Omega_m$ = 0.3 and $\Omega_\Lambda$ = 0.7.

\section{The  sample}

\begin{figure*}
\centering
\includegraphics[width=2.0\columnwidth]{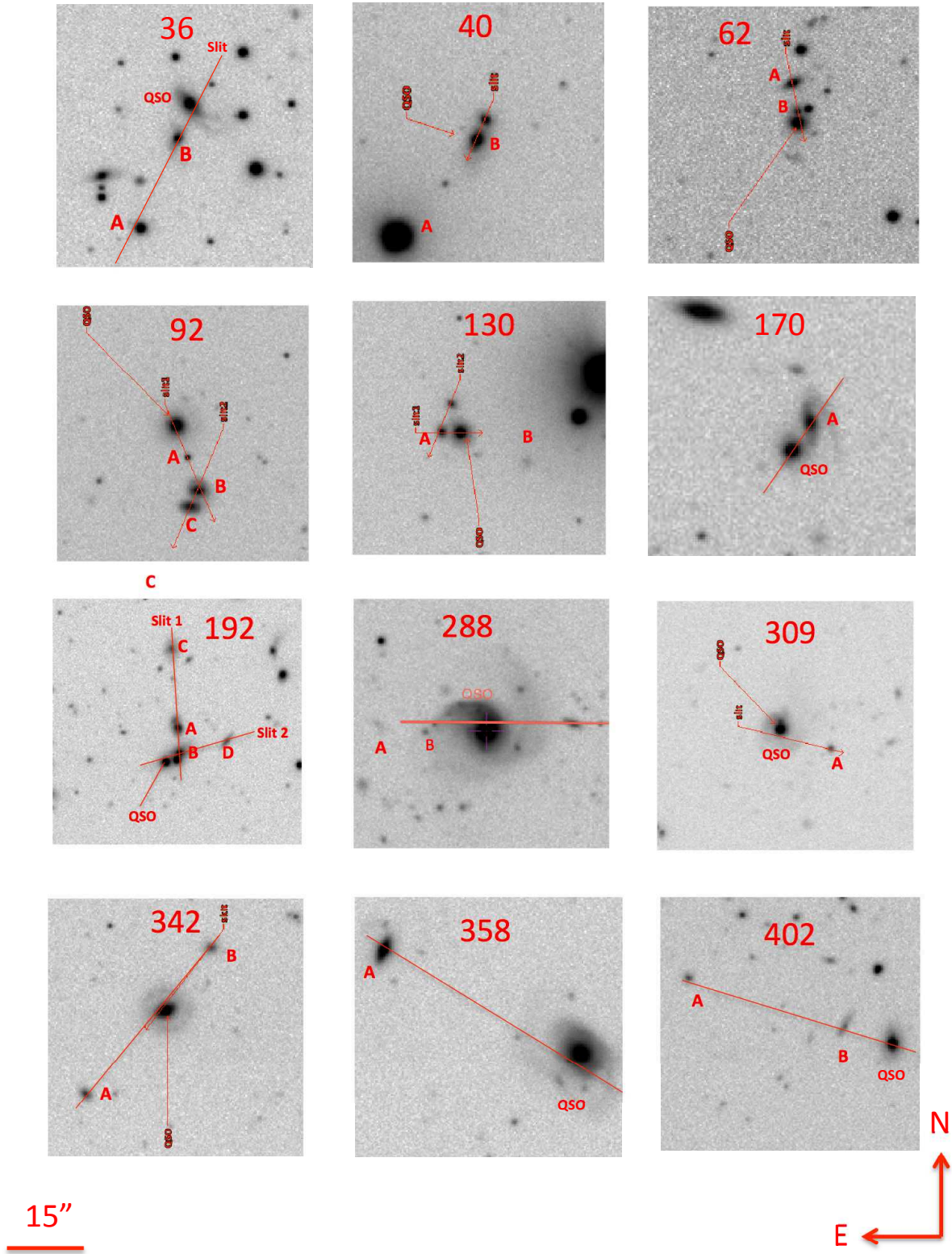}
\caption{SDSS images ($i$ filter) of the field around the  QSO showing the  companion galaxies for which spectra were secured (see also Table \ref{tab:sample} ) . The adopted slit positions are shown by a  red solid line.  }  

\label{fig:ima}
\end{figure*}

The targets are extracted from the sample of 52 low redshift (z $<$ 0.3) quasars for which we carried out a multicolor study of their host galaxies and close environments \citep{bettoni15}.
These targets belong to a larger dataset of $\sim$ 400 quasars  (z $<$ 0.5)    located in the SDSS-Stripe82 \citep{abazajian09,annis2011} that was extensively investigated for the environment and  host galaxy properties in our previous works \citep{falomo14,karhunen14,karhunen16,bettoni15}.  
We selected  18 QSOs with $z<0.3$  for which the host galaxy is well resolved \citep[see details in][]{falomo14} and for each objects of this sample there is at least one close (distance  $\leq 80$  kpc) companion galaxy with an apparent r band magnitude brighter than $r$=22. In order to test the possible association with the QSO we obtained  optical spectra of both the host galaxy and the close companion using the slit aligned along the two objects (see in Fig. \ref{fig:ima} the fields of view of our targets). Due to the observing conditions we could observe only 12 objects (i.e. 67\%) of the whole selected sample. 
In Table \ref{tab:sample} we give the list of the observed QSOs and their main properties.  For this sub-sample of QSO we obtained off-nuclear spectra of QSO hosts and of the close companions.

\section{Observations and data analysis}

The spectra of the QSO and of the close companions were collected with the Nordic Optical Telescope (NOT ) at La Palma. Observations were performed with grism \#7 on ALFOSC, this give a spectral resolution of R=500, with the used slit, 1.3 arcsecs wide. Our targets were observed in two different runs in Sept. 13-16 and Nov. 9-12 2015. For each QSO the slit was oriented slightly offset (1-2 arcsec corresponding to 4-10 kpc at the redshift of the targets) from the QSO nucleus and at a Position Angle (PA) that allows to take the spectra of the companion simultaneously. In Table \ref{tab:sample} for each QSO we give the observed PAs and offsets from the nucleus. 
\bigskip

Standard IRAF \footnote{IRAF is distributed by the National Optical Astronomy Observatory, which is operated by the Association of Universities for Research in Astronomy (AURA) under cooperative agreement with the National Science Foundation} tools were adopted for  the data reduction. Bias subtraction, flat field correction, image alignment and combination were performed. Cosmic rays were cleaned by combining different exposures with the {\sl crreject } algorithm. The spectra were then calibrated both in wavelength and in flux. The accuracy of the wavelength calibrations is  $\sim$0.2 \AA .
Since only a fraction of the flux from the companion galaxies is gathered through the slit we have set the absolute flux calibration of the spectra by normalizing the continuum to the flux corresponding to the r band magnitude 
of the galaxy as derived from the same SDSS images used in \citet{falomo14}. In the case of the spectra of the QSO host galaxies we instead normalized 
the flux of their spectra to the total magnitude of the host galaxies as derived in our previous study of the QSO hosts for the whole sample \citep{falomo14}.
The final spectra are presented in Figures \ref{fig:spec}.

We used the RVSAO IRAF Package to measure the redshift, both for emission (with {\it emsao})  and pure absorption spectra (with {\it xcsao}).  In the case of absorption line spectra we used as template a synthetic stellar spectrum (a KIII star) taken from the \citet{jacoby} library. In Table \ref{tab:res} we give our measured redshift.


\begin{figure*}
\includegraphics[width=2.0\columnwidth]{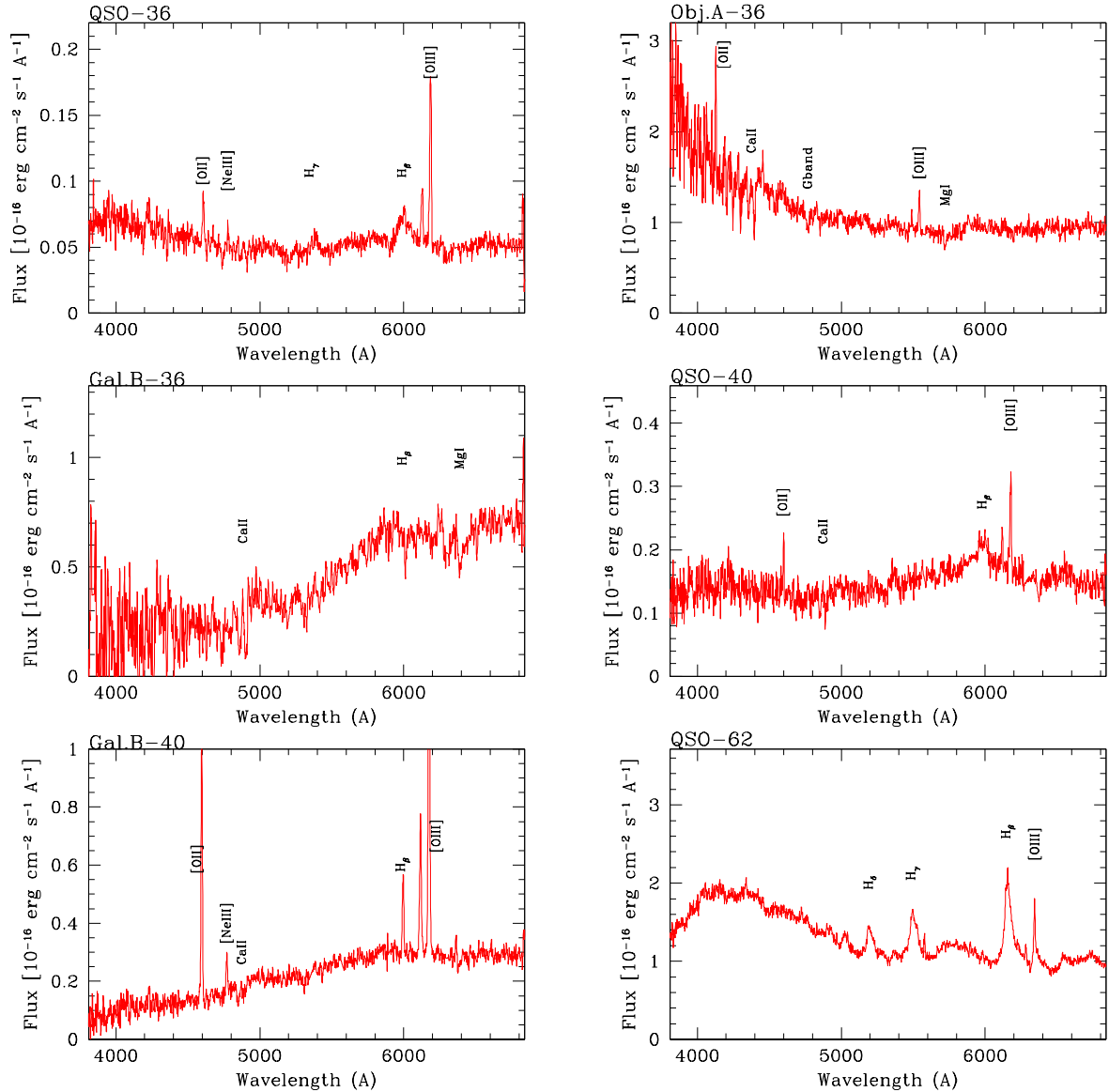}
\caption{Optical spectra of QSO and their close companion galaxies (see also \ref{tab:sample}  ) .  
Main emission and/or absorptions lines are marked by labels.  }
\label{fig:spec}
\end{figure*}

\begin{figure*}
\includegraphics[width=2.0\columnwidth]{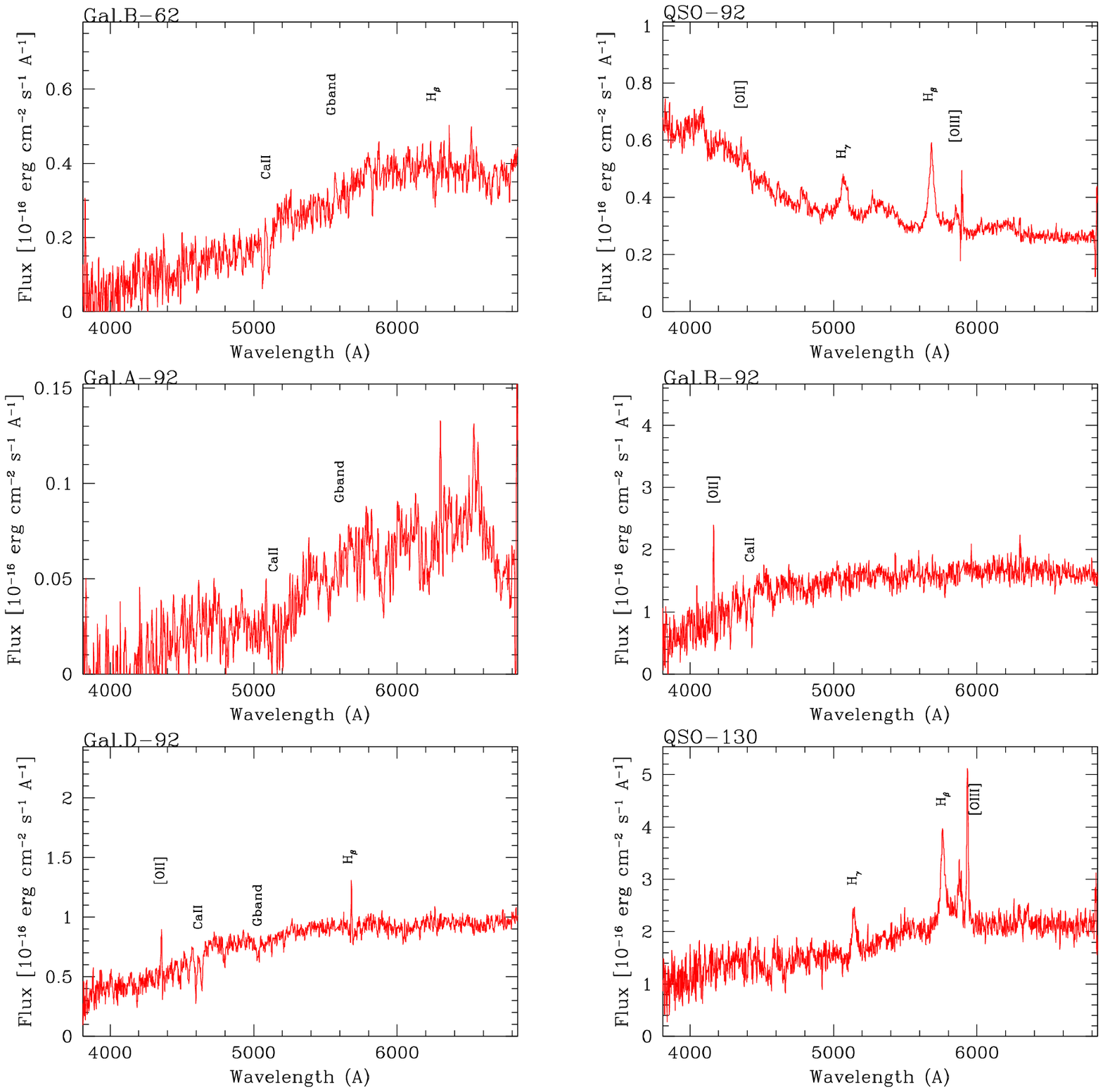}
\setcounter{figure}{1}
\caption{Optical spectra of QSO and their close companion galaxies, continue }
\label{fig:spec}
\end{figure*}

\begin{figure*}
\includegraphics[width=2.0\columnwidth]{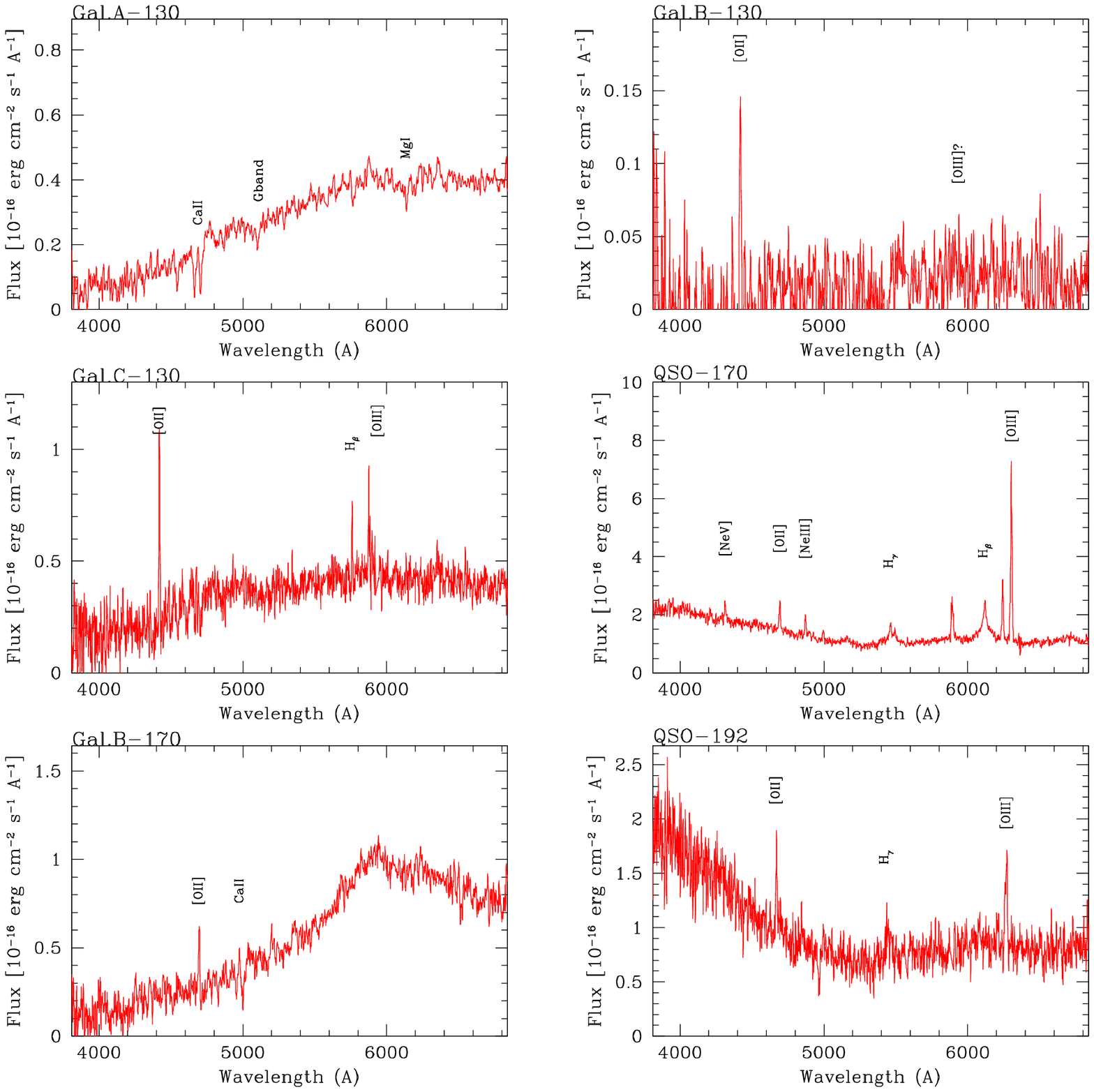}
\setcounter{figure}{1}
\caption{Optical spectra of QSO and their close companion galaxies, continue}
\label{fig:spec}
\end{figure*}

\begin{figure*}
\includegraphics[width=2.0\columnwidth]{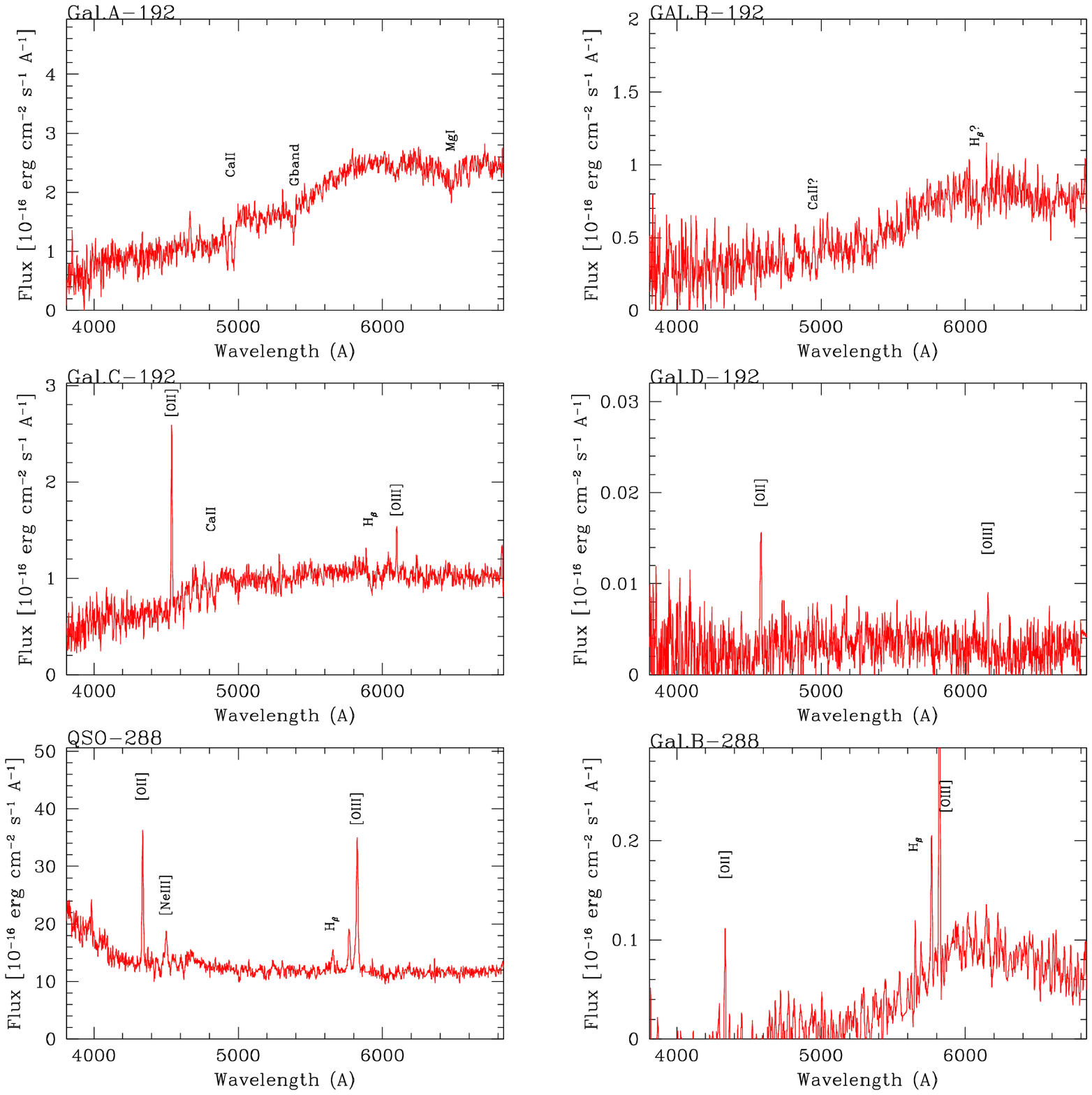}
\setcounter{figure}{1}
\caption{Optical spectra of QSO and their close companion galaxies , continue }
\label{fig:spec}
\end{figure*}

\begin{figure*}
\includegraphics[width=2.0\columnwidth]{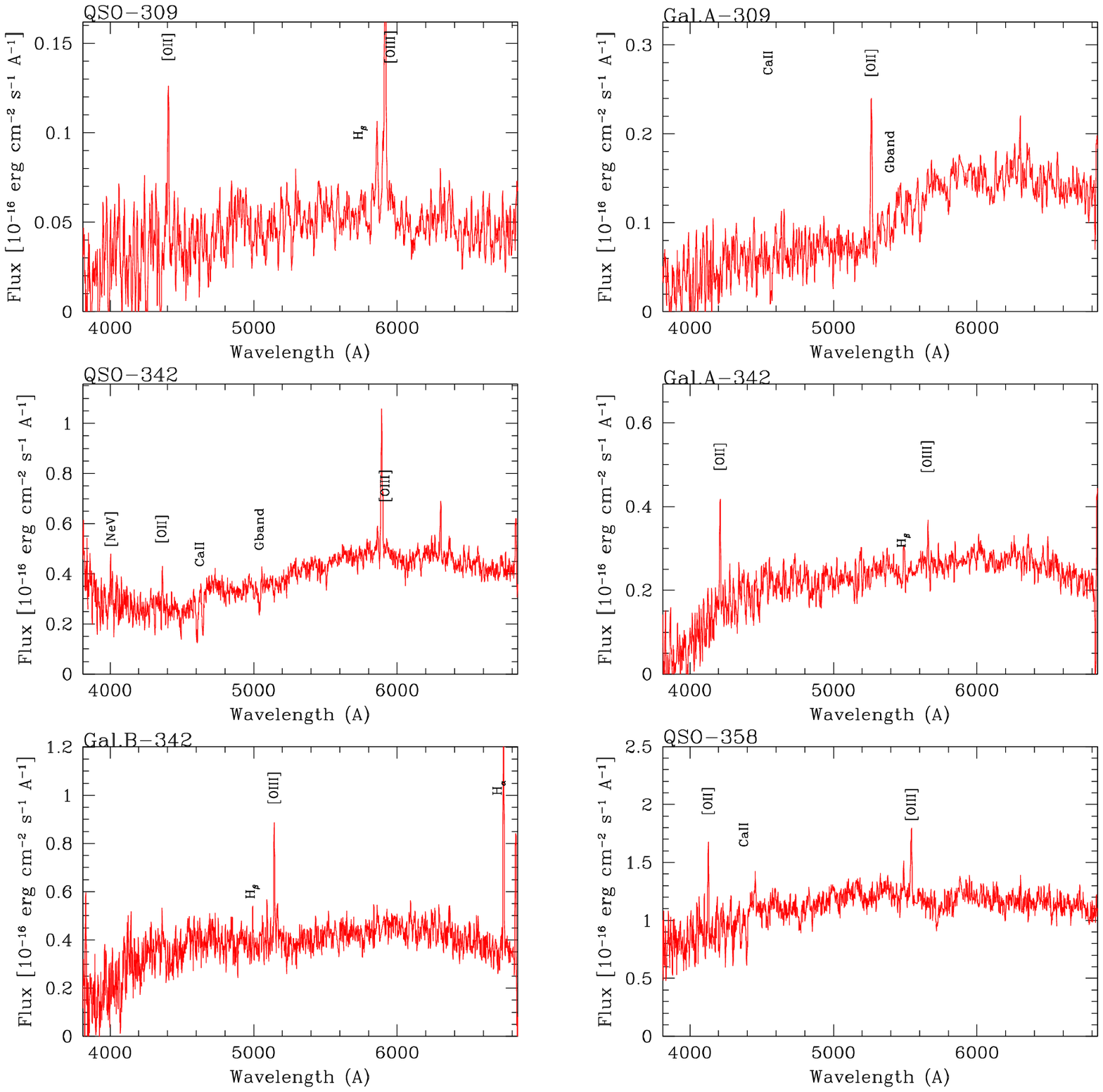}
\setcounter{figure}{1}
\caption{Optical spectra of QSO and their close companion galaxies,  continue }
\label{fig:spec}
\end{figure*}

\begin{figure*}
\includegraphics[width=2.0\columnwidth]{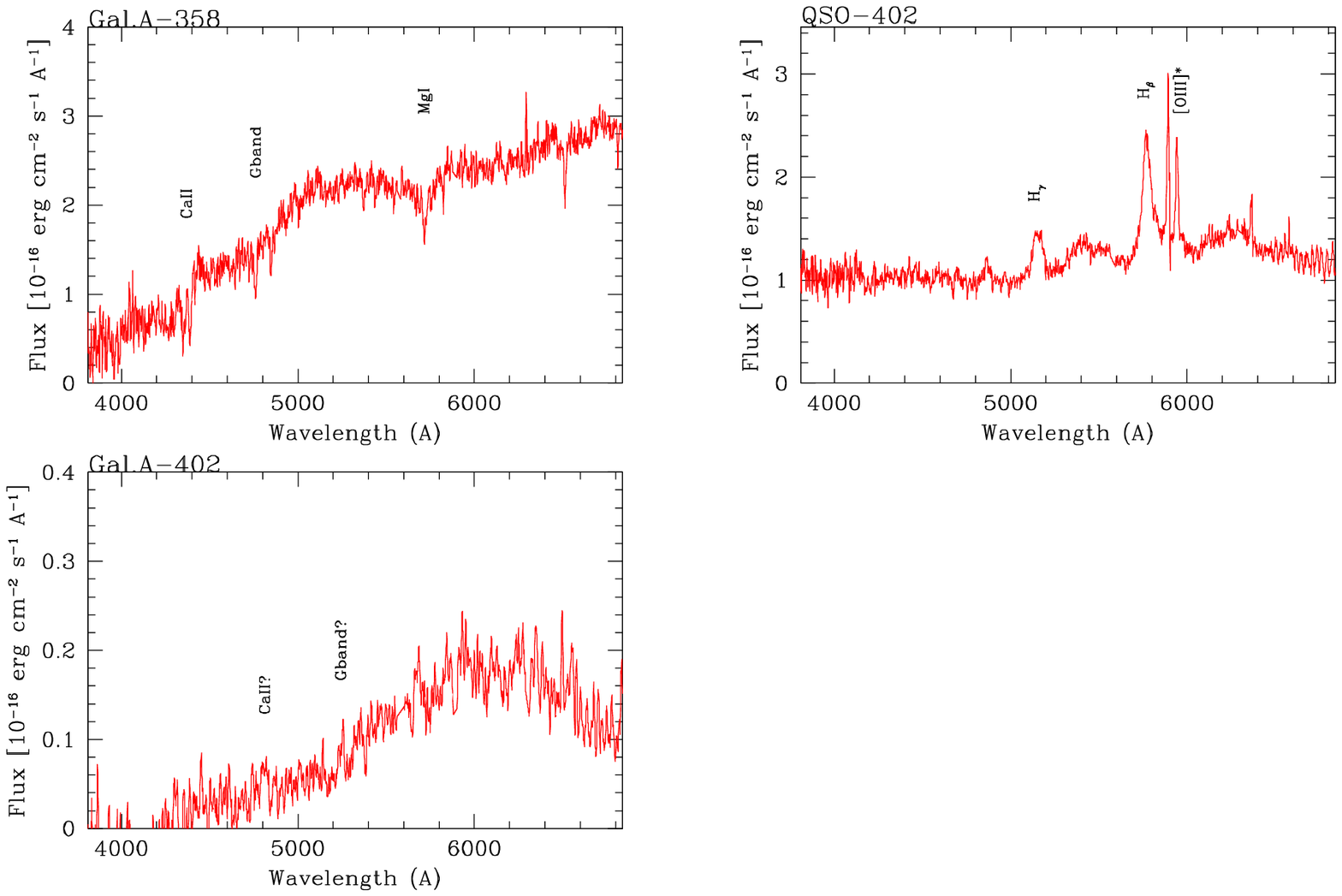}
\setcounter{figure}{1}
\vspace{-5truecm}
\caption{Optical spectra of QSO and their close companion galaxies,  continue }
\label{fig:spec}
\end{figure*}

 \begin{table*}
 \caption{The observed QSO and companion galaxies  }
\begin{tabular}{rclrrrrccccc}
\hline
  \multicolumn{1}{c}{Nr} &
  \multicolumn{1}{c}{Obj} &
  \multicolumn{1}{c}{SDSSJ} &
  \multicolumn{1}{c}{$z_{SDSS}$} &
  \multicolumn{1}{r}{PD} &
  \multicolumn{1}{c}{PA} &
  \multicolumn{1}{c}{offset} &
  \multicolumn{1}{c}{u} &
  \multicolumn{1}{c}{g} &
  \multicolumn{1}{c}{r} &
  \multicolumn{1}{c}{i} &
  \multicolumn{1}{c}{z} \\
& & & & kpc & deg & "~~~ & & & & & \\
\hline
  36 & QSO & 211234.88-005926.8 & 0.2351 & &  & 1.8 & 18.73 & 18.71 & 18.37 & 17.96 & 17.83\\
   & Obj. A & 211235.72-005958.9 &  & 130.1 & 155 & 2.5 & 19.42  & 18.22 & 17.75 & 17.58  & 17.49 \\
    & Gal. B & 211235.08-005935.7 &  & 53.1 & 155 & & 22.54 & 20.48 & 19.08 & 18.50 & 18.20\\
  40 & QSO & 211832.75+004500.8 & 0.2332 &  &  &1.35 & 18.69 & 18.60 & 18.34 & 17.94 & 17.91\\
   & Gal. B &211832.62+004505.4 &  & 27.9 &  150 & & 22.14 & 20.81 & 19.95 & 19.13 & 18.95\\
  62 & QSO & 215744.18+005303.6 & 0.2674 &  &  & &18.66 & 18.63 & 18.44 & 18.35 & 17.91\\
   & Gal. B & 215744.26+005313.8 &  & 67.4 & 10 & & 24.79 & 21.24 & 19.66 & 19.11 & 18.67\\
  92 & QSO & 222632.66-005717.7 & 0.1681 & &  &1.3 & 17.89 & 17.87 & 17.64 & 17.27 & 17.21\\
   &Gal. A & 222632.46-005726.3 &  & 35.5 & 25 & & 23.69 & 24.16 & 21.68 & 19.94 & 18.88\\
   &Gal. B & 222632.27-005735.3 &  & 72.2 & 150 & & 20.04 & 18.87 & 18.08 & 17.61 & 17.31\\
   &Gal. C & 222632.40-005739.8 &  & 87.3 & 150 & & 20.59 & 19.35 & 18.67 & 18.27 & 17.92\\
  130 & QSO & 231711.79-003603.6 & 0.1861 & &  & 1.0 & 17.81 & 17.97 & 17.82 & 17.40 & 17.48\\
   &Gal. A& 231712.12-003603.5 & & 21.7 & 12 & & 23.45 & 20.72 & 19.58 & 19.22 & 18.88\\
   &Gal. B& 231710.52-003603.8 &  & 82.5 & 12 & & 25.29 & 22.78 & 22.64 & 21.18 & 22.87\\
   &Gal. C& 231711.95-003555.8 &  & 35.5 & 152 & & 21.25 & 20.37 & 19.72 & 19.46 & 19.44\\
  170 & QSO & 000557.23+002837.7 & 0.2596 &  & & & 18.59 & 18.57 & 18.41 & 18.37 & 17.86\\
   &Gal. B& 000557.04+002842.2 &  & 34.2 & 160 &  & 21.55 & 19.84 & 18.67 & 18.11 & 17.76\\
  192 & QSO & 002831.71-000413.3 & 0.2519 &  &  & &18.33 & 17.99 & 17.67 & 17.38 & 17.11\\
    &Gal. A& 002831.40-000400.8 &  & 81.2 & 15 & & 20.11 & 19.32 & 18.58 & 18.14 & 17.95\\
   &Gal. B& 002831.36-000409.9 &  & 37.9 & 120 & & 22.57 & 20.80 & 18.50 & 18.04 & 17.57\\
    & Gal. C & 002831.57-000332.7 &  & 244.4  & 15 & & 21.18 & 20.20 & 19.44 & 19.11 & 19.12\\
     & Gal. D& J002830.16-000405.8 &  & 145.4 & 120 & & 26.38 & 22.11 & 20.44 & 19.87	 & 19.54\\
  288 & QSO. & 015950.24+002340.8 & 0.1627 &  &  & & 15.91 & 15.90 & 15.96 & 15.73 & 15.82\\
   &Gal. B& 015951.11+002342.8 & & 49.2 & 0& & 23.04 & 21.70 & 21.26 & 22.81 & 20.94\\
  309 & QSO & 021359.79+004226.7 & 0.1823 &  & & & 17.67 & 17.67 & 17.33 & 16.90 & 17.19\\
   &Gal. A& 021358.87+004221.3 &  & 62.72& 20 & & 23.49 & 21.87 & 20.63 & 20.16 & 19.70\\
  342 &QSO & 025334.57+000108.3 & 0.1705 &  &  &  1.3 &19.01 & 18.52 & 18.07 & 17.57 & 17.29\\
   &Gal. A& 025336.01+000045.6 & & 123.1 & 142 & & 21.86 & 20.60 & 20.01 & 19.59 & 19.71\\
   &Gal. B & 025333.76+000125.2 & & 82.3 & 142 & & 20.46 & 19.79 & 19.48 & 19.31 & 19.18\\
  358 &QSO & 030639.57+000343.1 & 0.1074 & &  & 2.3 &17.72 & 17.34 & 17.11 & 16.59 & 16.55\\
   &Gal. A& 030642.89+000409.6 &  & 132.5 & 65& & 20.65 & 18.64 & 17.63 & 17.18 & 16.78\\
  402 &QSO & 033651.52-001024.7 & 0.1868 & &  && 18.58 & 18.57 & 18.22 & 17.76 & 17.63\\
   &Gal. A&033652.31-001019.8 & & 56.2 & 65 & & 24.69 & 22.03 & 20.44 & 19.88 & 19.07\\

\hline
\hline\end{tabular}
\begin{list}{}{}
\item[]Column (1): identifier from \cite{falomo14}, column (2) the QSO galaxy identifier, columns (3) and (4) the name and redshift from SDSS. Column (5) PD the Projected Distance of the  companion in kpc, column (6) the observed Position Angle (PA),  column (7) the slit offset from the QSO nucleus, columns (8) to (12) the u, g, r, i, and z magnitudes from SDSS-DR7.
\end{list}
\label{tab:sample}
\end{table*}

%
\section{Results}

In Figure \ref{fig:spec} we report the optical  spectra of the 12 observed quasars and of  their companion galaxies. Results from the spectra of the companion objects are summarized in Table \ref{tab:res} and specific comments on individual objects given in the Appendix. In a number of cases there is more than one companion inside the slit thus in total we are able to secure  the spectra of 22 objects in the immediate environment around the targets. It turned out that 1 out of these 22 apparent companions is a star (object labeled A for QSO 40) and is not furthermore considered in this work (see Figure \ref{fig:ima}). Thus the final sample of analyzed spectra is composed of 21 galaxies.

\subsection{Close companion galaxies}

We found that for 8 QSO the selected companion galaxy at projected distance $<$ 80 kpc is associated with the quasars  (assuming $\Delta$V $<$ 400 km/sec) while in the remaining 4 cases the companions are either a foreground or a background galaxy.
 In one (\#192) out of the 8 cases where we have one associated companion galaxy there is another companion galaxy at projected distance  larger than 80 kpc that is at the same redshift of the QSO. In another one ( \#130) two more associated companion galaxies are found.  One of the three companion objects is a very faint galaxy ($M_i \sim$ -18.6), almost undetectable  in the broad band Stripe82 images but  with clear visible  [OII] and [OIII] emission lines in the spectrum (see Figure \ref{fig:spec}). The total number of companions turn out to be 11.
 
\begin{figure}
\centering
\includegraphics[width=1.0\columnwidth]{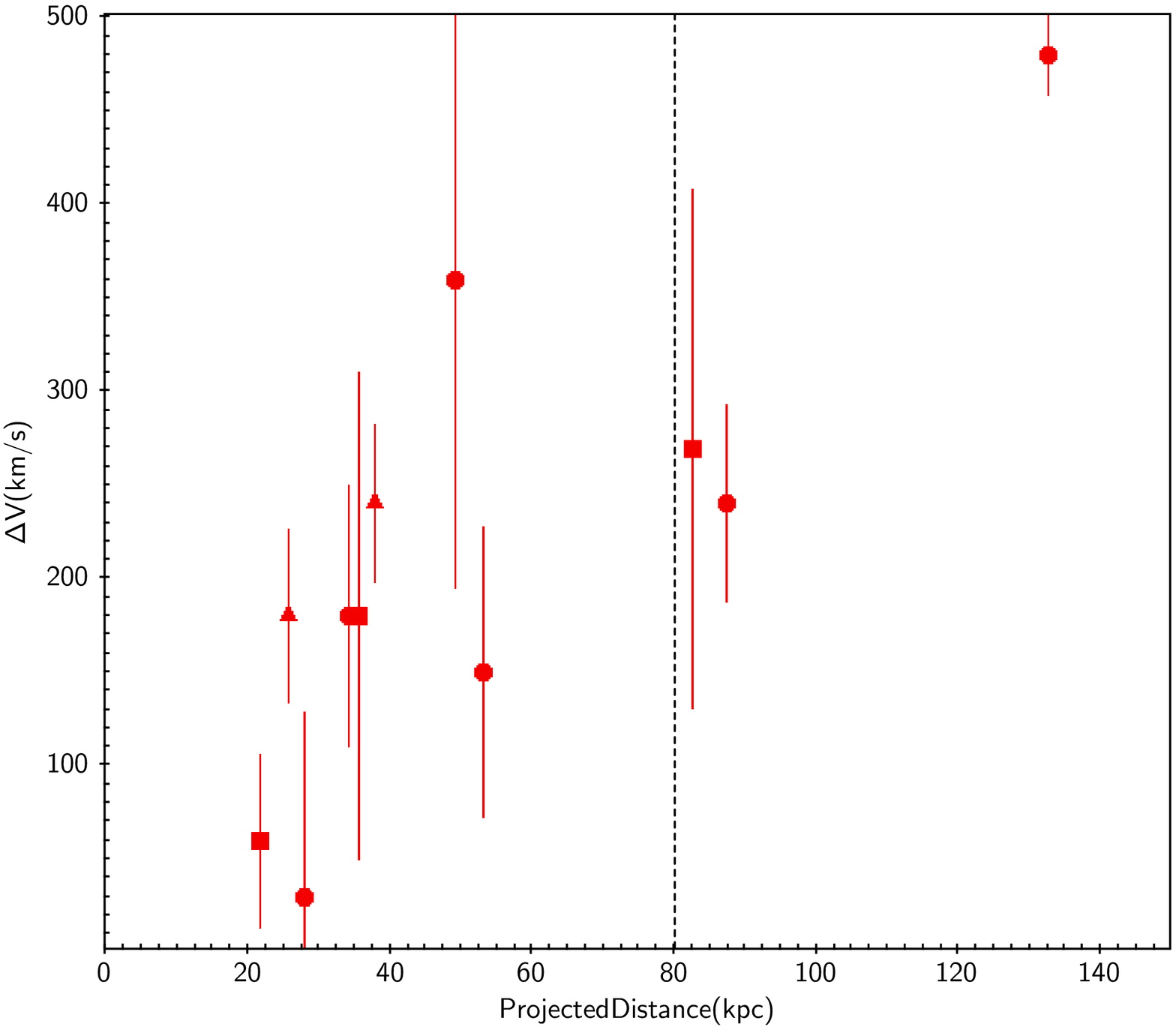}
\caption{The difference of radial velocity between the QSO and the associated companion galaxy  versus the projected distance of the companion. Different symbols refer to QSO with only one companion galaxy (filled circles), with two companion galaxies (filled triangles) and with three companions (filled squares) . The dashed vertical line indicates the projected distance of 80 kpc (see text).
}
\label{fig:redshift}
\end{figure}

 In fig \ref{fig:redshift} we show the distribution of $\Delta V$ (the difference of radial velocity between the companion galaxy and the QSO) with respect to the projected distance of the associated companion galaxies. Closer companions are found to have also the smallest $\Delta V$.

\subsection{Star formation from [OII] lines}

 \begin{table*}
 \caption{[O II] emission line measurements}
\begin{tabular}{llcccccc}
\hline
  \multicolumn{1}{c}{Obj} &
  \multicolumn{1}{c}{$z_{our}$} &
  \multicolumn{1}{c}{log(Flux)} &
  \multicolumn{1}{c}{$log(L)$} &
    \multicolumn{1}{c}{$SFR$}&
        \multicolumn{1}{c}{$sSFR$}&
        \multicolumn{1}{c}{$log(M^*)$}\\

&& erg $s^{-1}$ & $L/L_{\odot}$ & $M_{\odot}$/yr  & $yr^{-1}\times10^{-10}$ & $M/M_{\odot}$\\
  \hline
  QSO-36     &   0.2355     &  39.86$\pm$0.09   &   6.27 & 0.5$\pm$0.1 & 0.1 &10.66 \\
  A     &      0.1070  &  40.40$\pm$0.06   &      6.82  & 0.5$\pm$0.05 & 0.5& 10.04 \\     
    B     &      0.2356  &  --   &    --  &  -- & -- & --\\    
 QSO-40     &   0.2338     &  40.03$\pm$0.05   & 6.44 & 0.6$\pm$0.06  & 0.1& 10.54\\
  B     &     0.2333  &  41.22$\pm$0.06  &    7.63 & 6.9$\pm$0.07 & 2.9 & 10.38   \\
 QSO-62      &   0.2656     &  --   &    --  &  -- & -- & --\\
  A     &      0.2873  &  --   &    -- & -- & -- & -- \\
 QSO-92      &   0.1686     &  39.70$\pm$0.09   &    6.11 &  0.3$\pm$0.06  & 0.1 & 10.54  \\
  A     &     0.3141   &  --   &   -- & -- & -- & --	\\
  B     &   0.1175 &  40.54$\pm$0.05   &    6.95 & 1.5$\pm$0.15  & 0.6 & 10.38	\\
  C     &      0.1687  &  --   &    -- & --	& --  & -- \\
 QSO-130      &   0.1853     &   --   &   -- & -- & -- & -- \\
  A     &      0.1863  &     --   &   -- & -- & -- & -- \\
  B     &      0.1852  &   40.28$\pm$0.35   &    6.70 & 0.1$\pm$0.01  & 20.3& 7.9  \\
  C     &     0.1855  &  40.82$\pm$0.06  &      7.24  &  1.0$\pm$0.1 & 1.5& 9.8 \\
 QSO-170      &   0.2590    &  41.37$\pm$0.05   &     7.78 & 14.3$\pm$1.4 & 3.3 & 10.64	\\
  B     &     0.2590   &  40.89$\pm$0.06   &      7.30& 6.7$\pm$0.7 & 0.5 & 11.13\\
  QSO-192    &   0.2526     & 41.26$\pm$0.05   &  7.67 & 15.3$\pm$1.5  & 1.3 & 11.06 \\
  A     &      0.2178  &  41.37$\pm$0.05   &   7.78 & 15.4$\pm$1.5   & 3.0& 10.71 \\
  B     &      0.2511  &  --   &     -- & --	& -- & -- \\
  C     &      0.2295  &   38.92$\pm$0.11  &   5.34   & 0.02$\pm$0.01 & 0.01&10.15	\\
  D    &      0.2513  &  -- &   -- & -- & -- & --	\\
  QSO-288     &   0.1636    &  42.27$\pm$0.05  &     8.69 & 143.7$\pm$14.1 & 18.1 & 10.9	\\
  B     &      0.1639  & 40.01$\pm$0.12   &    6.43 & 0.8$\pm$0.08 & 0.1 & 10.93	\\
 QSO-309      &   0.182     & 40.07$\pm$0.05  & 6.48 & 0.8$\pm$0.08  & 0.1 & 10.71 \\
  A     &      0.4124  & 40.95$\pm$0.07  &  7.37  & 6.1$\pm$0.06  & 1.1 & 10.76  	\\
 QSO-342      &   0.1691     &  40.02$\pm$0.05  & 6.43 & 0.7$\pm$0.07 & 0.1 & 10.69	\\
  A      &    0.1297   &  40.03$\pm$0.07  & 6.45 & 0.1$\pm$0.01 & 0.6& 9.25	\\
  B      &    0.0273   & -- &  -- & -- & --	& -- \\
 QSO-358        &  0.1070      & 40.20$\pm$0.05 &  6.61& 0.7$\pm$0.07 & 0.3 & 10.43        \\
  A        &0.1058 & --  &   -- & -- & -- & -- \\
 QSO-402       &  0.1869    & --    &   -- & -- & --	& -- \\
  A        & 0.2322 & --    &  -- & --  & -- & --	\\
\hline\end{tabular}
\begin{list}{}{}
\item[]Column (1): identifier from \cite{falomo14}, column (2) the redshift from our measurements. Column (3) log of the flux of [OII] line (error is obtained combining the uncertainity in the flux calibration based on SDSS-DR7 photometry and the measurement of line fluxes); 
column (4) log of the luminosity; column (4) the derived SFR in $M_{\odot}$/yr , column (5) specific SFR in $yr^{-1}$ and column (6) the stellar mass in $M_{\odot}$.
\end{list}
\label{tab:res}
\end{table*}

For 20 (9 QSOs and 11 galaxies in the fields) out of 33 observed sources we detect a significant emission line of [OII] 3727 \AA. 
Only 5 of the 11 galaxies with [OII] 3727 emission are associated to the QSO  and are used in our analysis but for completeness in Table \ref{tab:res} we report the values for the non associated  companions for comparison.
The intensity of this line can be used as an approximate tracer of the  star formation rate (SFR) \citep[e.g.][]{Gal89,K98} because the relationship between [OII] luminosity and SFR may be  affected by reddening and relative abundance. The observed [OII] luminosity ranges from 10$^{39}$ to 2 x 10$^{41}$ erg s$^{-1}$ with the exception of QSO 288 that exhibits an [OII] luminosity $>$ 10$^{42}$ erg s$^{-1}$. 
On average these luminosities cover the faint tail of the L$_{[OII]}$ luminosity of QSO \citep{kalfountzou12}.
In table \ref{tab:res} we report the intensity and the luminosity of the measured [OII] emission lines for all observed galaxies.
To estimate the SFR we adopt the recipe proposed by \cite{gilbank10} (their equation 8; see also \citet{gilbank11} ) that takes in to account,  
in an empirical form, the systematic effects of mass (metallicity) for the SFR vs [OII] luminosity  relationship. 
The mass of the galaxies for which we secure the spectra was estimated using the 
empirical relationship between stellar mass, absolute magnitude and color derived by \citet{kauffmann03} and \citet{gilbank10} from multicolor SDSS imaging of a very large sample of low redshift (z $\sim$ 0.1) galaxies. 
The stellar masses of our galaxies were derived using the SDSS magnitudes and applying k-correction using the routine KCORRECT  \citep{blanton07}  for consistency with the derived empirical relation. The values for the masses and SFRs are given in Table \ref{tab:res}. Since we cannot 
estimate directly the intrinsic reddening for the observed galaxies from the spectra (e.g. from Balmer decrement), 
to compute the SFR from \citet{gilbank10} (eq.  8) assuming  a nominal factor of 2 for the ratio of extinction between [OII] and  H$_{\alpha}$  \citep[e.g.][]{K98}. 
The adopted relation takes into account the dependence of SFR from dust extinction and metallicity using an empirical relation with the stellar mass   \citep[see][]{gilbank10}. 
We find that the  SFR for all observed companion galaxies is in the range  $\sim$0.02 to $\sim$15 M$_\odot yr^{-1}$. For 5 companion galaxies associated to the QSO the median SFR is 1.0 $\pm$ 0.8  (error represents the semi inter-quartile range) while for 6 that are  not associated it is 1.5 $\pm$ 0.1. 
For the QSO host galaxies in nine cases we detect [OII] emission. For five of them the SFR is  modest ($<$ 1 M$_\odot yr^{-1}$) while in two  cases (QSOs 170 and 192) an high SFR was found ($\sim$ 10 $M_\odot yr^{-1}$) and in one case (QSO 288) we estimate
a very high SFR ($\sim$ 140 $M_\odot yr^{-1}$). The specific SFR (sSFR = SFR/M$_*$) for most of the galaxies in which [OII] emission is detected is much smaller than 1 Gyr$^{-1}$. This is suggestive of only modest star formation integrated over the Hubble time. Only in few cases a greater sSFR is found that could be due to a recent episode of star formation.

 We note that the values of SFR as  derived from the [OII] luminosity assume that the observed L$_{[OII]}$ luminosity is representative of the whole galaxy. This is a reasonable assumption for most of the companion galaxies because 
 their size is comparable to or somewhat larger than the slit aperture used for the spectra. However, in some cases (e.g the QSO 288 host; see Figure \ref{fig:ima}) the slit aperture intercepts only a fraction of the 
 host galaxy, therefore, any gradient or structured emission may bias this measurement. 
 \begin{figure}
\includegraphics[width=1.7\columnwidth]{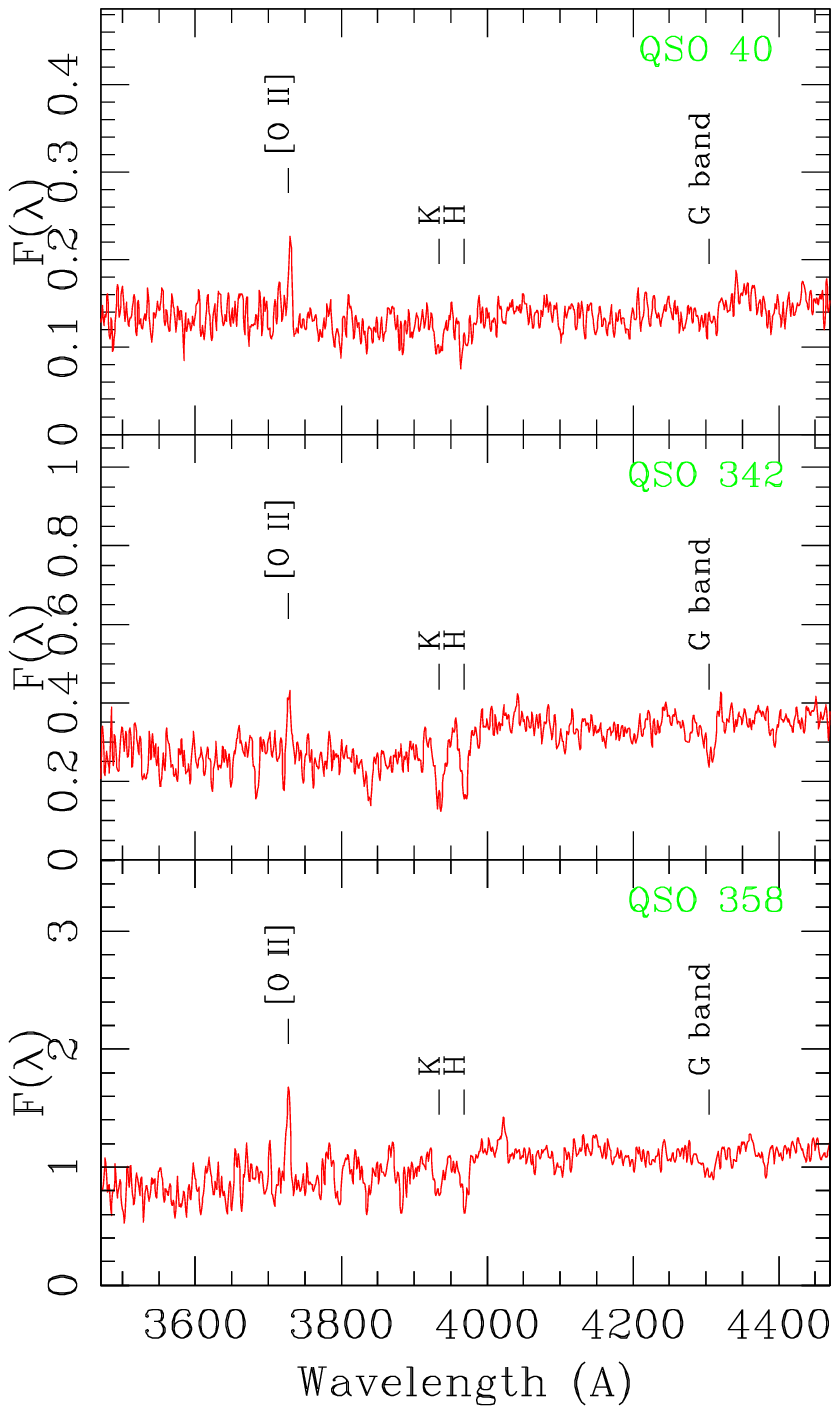}
\vspace{-4truecm}
\caption{The rest frame spectra of the QSO host galaxies for targets \# 40, \# 342 and \# 358. Clear signature of the underlying stellar population is apparent. In all cases a significant [OII] 3727 \AA emission line is detected. The spectra were obtained through a slit that was offset from the nucleus (see details in Table \ref{tab:sample}).}
\label{fig:ima_spec}
\end{figure}

\subsection{Spectra of the host galaxies}

In addition to the optical spectra of the companion galaxies we secured off slit ($\sim$ 1 -2 arcsec) spectra of the QSO in order to better gather the starlight signal from their host galaxies see Figure \ref{fig:ima} and \ref{fig:spec}. In the cases where more than one companion could be observed in the same slit the orientation was set to detect two companions (see Table \ref{tab:sample}) 

Depending on the brightness of the nucleus and on the position of the slit some  spectra of the targets (QSO)  contain therefore the flux contribution from a fraction of the nucleus light and from the host galaxy. 
For QSO \# 40, 342,  358 we are able to detect Ca II H, K and G band absorption features from the old stellar population of the QSO hosts (see Figure \ref{fig:ima_spec}). In all these cases [OII] 3727 \AA~ emission is also present. Because to the strong emission from the QSO some contribution of the [OII] flux could be due to the nucleus. However, since [OII] originates mainly in HII regions the contamination from the nucleus should be unimportant. Note that in 2 out of these 3 cases  there is also a close companion galaxy associated to the QSO. 

\section{Summary and Conclusions}

We obtained optical spectroscopy of 21 close companions galaxies for  12 low redshift (z $<$ 0.3) QSOs. 
It turns out that 11 of them are associated to the QSO and most of them are at projected distance $<$ 80 kpc. 
In two cases we found that more than one companion galaxy is associated with the QSO.
In 5 cases of associated companions their spectra exhibit significant [OII] 3727 \AA~ emission. These companions are at projected distances of 20-50 kpc therefore if the ionization source is due to the QSO  the companion galaxy occupies a small solid angle as seen from the quasar, and  only a small fraction of the ionizing flux will hit the companion. Under these conditions the [OII] emission line intensity is directly related to the ionization parameter \cite[c.f.][]{gnedin97} that is significantly reduced by the covering factor ($\sim$ 10$^{-3}$ for a companion of 3 kpc at a distance of 40 kpc). The EW of [OII] lines of these companion galaxies cover a wide range (3 to 80 \AA) and are comparable to the few other spectroscopic measurements of close companion galaxies reported by \cite{gnedin97}.
The average level of SF (as derived from [OII] luminosity) of companion galaxies that are associated to the QSO appears similar to that of the companion galaxies that are not associated to  the QSO.  The majority (9 out of 11) of the observed QSO exhibit [OII] emission, however, only in three cases the SFR is 
significant.  Two of them (QSO \#170 and \#288)  have also associated close companions with [OII] emission while in one case (QSO \#192) no [OII] emission in associated companions are found. 
These results  suggest (albeit still based on a  scanty statistics) a modest role of the QSO emission for the SF in nearby companion galaxies. 

For three objects we are also able to detect the starlight spectrum of the QSO host galaxy the spectrum of which is characterized by prominent absorption lines of old stellar population and low level of SFRs  \citep[c.f. also][]{matsuoka15} . 

\section*{Acknowledgments}

We thanks the anonymous referee whose comments improved the paper. Funding for the SDSS and SDSS-II has been provided by the Alfred P. Sloan Foundation, the Participating Institutions, the
National Science Foundation, the U.S. Department of Energy, the National Aeronautics and Space Administration, the Japanese Monbukagakusho, the Max Planck Society, and the Higher Education Funding Council for England. The SDSS Web Site is http://www.sdss.org/.

The SDSS is managed by the Astrophysical Research Consortium for the Participating Institutions. The Participating
Institutions are the American Museum of Natural History, Astrophysical Institute Potsdam, University of Basel, University
of Cambridge, Case Western Reserve University, University of Chicago, Drexel University, Fermilab, the Institute for
Advanced Study, the Japan Participation Group, Johns Hopkins University, the Joint Institute for Nuclear Astrophysics, the Kavli Institute for Particle Astrophysics and Cosmology, the Korean Scientist Group, the Chinese Academy of Sciences (
LAMOST), Los Alamos National Laboratory, the Max-Planck-Institute for Astronomy (MPIA), the Max-Planck-Institute for Astrophysics (MPA), New Mexico State University, Ohio State University, University of Pittsburgh, University of
Portsmouth, Princeton University, the United States Naval Observatory, and the University of Washington.

\section{Appendix - Notes to individual objects}

\bigskip
{\bf QSO 36 }
\bigskip

Radio quiet quasar at z = 0.2356 as derived from [OIII] emission lines. The optical spectrum of closest companion galaxy (labeled B, r = 19.1; dist. 53 kpc from QSO) shows clear absorption features of Ca II, G band, H$_\beta$, and MgI 5175\AA ~at z = 0.2356. 
The chosen slit position intercepts also  another compact object (label A) at 33.5 arcsecs from the QSO. The slit position was only barely tangent to this object that is classified as star by SDSS. A very faint diffuse companion is present at $\sim$ 3 arcsecs from object A and intercepted by the slit. Our spectrum is characterized by a blue continuum with emission lines of [OIII] and [OII] and stellar absorption features (CaII, G band, Na I)  at the redshift z=0.107.

\bigskip
{\bf QSO 40 }
\bigskip

This QSO (z = 0.2335 ) has a very close companion galaxy (labeled B, r =  20.0 ) at 27 kpc projected distance (PD). Our optical spectrum shows prominent (EW = 63 \AA) [OIII] 5007\AA ~emission line and also very strong [OII] 3727 \AA ~(EW = 71 \AA). In addition we also detect emission of [NeIII] 3869\AA ~and H$_\beta$ and Ca II absorptions.
All these lines correspond to the redshift z = 0.2332 that is identical within the measurement errors ($\Delta$V = 150 km/s) to that of the QSO. 
This is thus a clear case of physical association QSO-close companion galaxy with evidence of significant star formation. The slit crosses also object labeled A that is a foreground star.

\bigskip
{\bf QSO 62}
\bigskip

The QSO (z= 0.2674) has a quasi  edge on  spiral galaxy (labeled A, r = 19.7) at PD = 67 kpc. The spectrum of this companion is characterized only by absorption features (Ca II, G band, H$_\beta$) at z = 0.2868. The large ($\sim$ 6000 km/s) difference of radial velocity excludes a present  association with the quasar.
At $\sim$ 2.5 arcsecs (corresponding to PD = 16 kpc) from the QSO there is a fainter (r = 21.1) compact companion (labeled B, see Figure \ref{fig:ima}) that is only marginally detected by our spectrum.

\bigskip
{\bf QSO 92}
\bigskip

In the immediate environment of this QSO (z = 0.1681) there are two similar galaxies at PD = 72  kpc (B; r= 18.1) and PD = 87 kpc (C r = 18.7). 
Our spectra shows that only galaxy C can be associated to the QSO having a difference of radial velocity of $\sim$ 230 km/s. The other companion galaxy (B) is a foreground object (z = 0.1175). Another fainter and closer compact companion (A; see Figure \ref{fig:ima}) is instead a background object at z $\sim$ 0.3.
Also in this case the associated companion galaxy (C) in addition to absorption lines from the stellar population of the galaxy shows also moderate [OII] 3727\AA ~emission line (EW = 4.6 \AA). 

\bigskip
{\bf QSO 130}
\bigskip

There are two close companions (A,C) at PD of 22 and 35 kpc, respectively. Both companions are at the same redshift ($\Delta$V $\sim$  50 km/s ) as the QSO (z = 0.1861).
While companion A has a pure absorption spectrum, the companion galaxy C shows a prominent [OII 3727] emission (EW = 30 \AA ) and also [OIII]  5007 \AA~  and H$_\beta$ 
(see Figure \ref{fig:ima}). Curiously we found another emission line galaxy at the same redshift at $\sim$ 19.5 arcsecs W of the QSO (PD = 82 kpc) in the spectrum obtained with the slit oriented along the QSO-gal. A direction (E-W). 

\bigskip
{\bf QSO 170}
\bigskip

The optical spectrum of the spiral companion galaxy (r = 19.8; PD = 34 kpc) to this QSO (z= 0.2596) shows both absorption lines (Ca II) and again [OII] 3727\AA ~emission at z = 0.2595 ($\Delta$V $<$  100 km/s  with respect to the QSO). 

\bigskip
{\bf QSO 192}
\bigskip

The immediate environment around this QSO (z= 0.2519 ) is rather complex (see Figure \ref{fig:ima}). There are two close companions the closest at $\sim$ 3.5 arcsecs W is a star and the other is a galaxy (labeled B) at $\sim$6 arcsecs NW at a redshift very close to that of the QSO (z=0.2511). We observed this field at two PA (15$^{\circ}$ and 120$^{\circ}$) obtaining the spectra of 3 more close objects (A, C and D) all are galaxies, A and C are foreground objects, while D has a redshift close to the QSO (z= 0.2513).


\bigskip
{\bf QSO 288}
\bigskip

This QSO (z=0.1627) is hosted by a face-on spiral galaxy. The optical spectrum of the closest (object B, PD $\sim$ 50 kpc) faint  (r= 21.3) companion turned out to be at z=0.1627.

\bigskip
{\bf QSO 309}
\bigskip

The faint (r=20.6) companion galaxy of this QSO (z= 0.1823) at PD = 63 kpc (object A) is a background emission line galaxy at z = 0.4129. The host galaxy shows a faint spiral structure.

\bigskip
{\bf QSO 342}
\bigskip

There are two companion galaxies (see Figure \ref{fig:ima}) that encompass this QSO (z=0.1705) at PD = 123 (A) and PD= 82 (B) . Either these companions are foreground emission line galaxies at z = 0.1297  and z = 0.0273, respectively. Our optical spectrum was secured along the direction (PA = $142^{\circ}$) connecting the companion galaxies A and B (see Figure \ref{fig:ima}). With such position we are also able to intersect the flux of the host galaxy of the quasar (at 1.3 arcsecs from its center). The spectrum of the QSO host (see Figure \ref{fig:ima}). 
We detect stellar absorption lines as H,K of Ca II, G band, MgI 5875 \AA, Ca+Fe.

\bigskip
{\bf QSO 358}
\bigskip

QSO (z=0.1074)  hosted by a spiral galaxy with a tidal tail. We tool the spectrum of the relatively bright (r=17.6) edge-on companion galaxy that is at 133 kpc from the QSO in the direction of the extension of the tidal tail (see Figure \ref{fig:ima}). 
We found a pure absorption line spectrum for this companion that is at z=0.1058. The difference in terms of radial velocity between the companion and the QSO is $\sim$ 500 km/s thus  it is unlikely that they form a bound system but  suggests that there was some interaction in the past.

\bigskip
{\bf QSO 402}
\bigskip

We took the spectrum of the faint (r=20.4) companion object at PD= 56 kpc that turned out to be at z = 0.2322. 

%

\end{document}